\documentclass[12pt]{iopart}
\usepackage{iopams}
\usepackage{graphicx}
\usepackage[usenames,dvipsnames]{color}
\usepackage{hyperref}
\usepackage[square,sort,comma,numbers,compress]{natbib}
\hypersetup{
     colorlinks=true,
     citecolor=blue,
     linkcolor=blue,
     urlcolor=blue}

\begin{document}
\title[]{Geometry-induced modification of fluctuation spectrum in
quasi-two-dimensional condensates}
\author{Arko Roy$^{1,}$$^2$\footnote{Corresponding
author: arkoroy@prl.res.in} and D. Angom
$^1$\footnote{Corresponding author: angom@prl.res.in}}
\address{$^1$Physical Research Laboratory,
             Navrangpura, Ahmedabad-380009, Gujarat,
             India}
\address{$^2$Indian Institute of Technology,
             Gandhinagar, Ahmedabad-382424, Gujarat, India}

\ead{angom@prl.res.in}


\begin{abstract}
 We report the structural transformation of the low-lying spectral modes,
especially the Kohn mode, from radial to circular topology as harmonic
confining potential is modified to a toroidal one, and this corresponds to a
transition from simply to multiply connected geometry. For this we employ
the Hartree-Fock-Bogoliubov theory to examine the evolution of low energy
quasiparticles. We, then, use the Hartree-Fock-Bogoliubov theory with the
Popov approximation to demonstrate the two striking features of quantum and
thermal fluctuations. At $T=0$, the non-condensate density due to interaction
induced quantum fluctuations increases with the transformation from
pancake to toroidal geometry. The other feature is, there is a marked change
in the density profile of the non-condensate density at finite temperatures
with the modification of trapping potential. In particular, the
condensate and non-condensate density distributions have overlapping
maxima in the toroidal condensate, which is in stark contrast to the case of
pancake geometry. The genesis of this difference lies in the nature of the
thermal fluctuations.
\end{abstract}

\vspace{2pc}

\maketitle


\section{Introduction}
 Topological defects are generated spontaneously during phase transitions,
and Bose-Einstein condensation is no exception. One approach to understand
the seeding of the topological defects is through the Kibble-Zurek (KZ)
mechanism \citep{kibble_76,zurek_85}. In recent times, the realization of 
Bose-Einstein condensates (BECs) in multiply connected (toroidal) geometries
\cite{heathcote_08,ramanathan_11,moulder_12,marti_15} have opened up new 
vistas to examine this mechanism in detail. Here, one may ask ``What is the 
basic nature of the fluctuations in toroidal condensates ?" Answer to this 
question can provide fundamental understanding on the scale, structure, and 
energetics of the defect formation. It is well known that 
dimensionality, and fluctuations play a major role in the formation 
of BECs. In brief, a three-dimensional (3D) Bose gas can undergo a phase 
transition to a BEC at non-zero temperatures. However, following the 
Mermin-Wagner-Hohenberg theorem~\cite{hohenberg_65,mermin_66,hohenberg_67}, 
thermal fluctuations restrict the transition to BEC for a two-dimensional (2D) 
Bose gas. The 2D Bose gases instead undergo a 
Berezinskii-Kosterlitz-Thouless
(BKT)~\cite{berezinskii_72,kosterlitz_72,kosterlitz_73} transition to a 
quasicoherent superfluid state~\cite{stock_05,hadzibabic_06,schweikhard_07},
and similar transition occurs occurs in quasi-2D Bose systems
~\cite{hadzibabic_06,schweikhard_07,clade_09,tung_10,hung_11,yefsah_11}. 
The BKT transition is relevant to the simply connected geometry (pancake)
of the system under consideration in the present work. We, however, focus 
our attention to the toroidal geometry as the effects of thermal fluctuations 
in KZ mechanism is our main interest, and use the pancake geometry
as the initial state for comparison.

 Apart from phase transitions, topological defects are also created in
fluid flows as well. 
In BECs, the velocity of pointlike defects, which must exceed a critical 
value, plays a vital role in the nucleation of topological defects like 
vortices. In simply connected BECs the critical velocity to 
create excitations due to a pointlike defect or a non-stationary obstacle has 
been experimentally 
measured ~\cite{raman_99,onofrio_00,engels_07,neely_10}, and similar 
study with toroidal BECs were reported in a recent experimental 
work~\cite{wright_13}. Furthermore, persistence and decay of superfluid 
flow in toroidal BECs have been observed in experiments through the 
appearance of vortex-antivortex pairs~\cite{ramanathan_11} in the presence 
of an obstacle exceeding a critical velocity. 

 To gain insights on the quantum, and thermal fluctuations in toroidal 
condensates vis-a-vis pancake shaped condensates, we examine the quasiparticle 
spectrum as the confining potential is modified from harmonic to a Mexican hat 
potential. We demonstrate, using the Hartree-Fock-Bogoliubov theory, novel 
features in the evolution of the low-energy quasiparticle amplitudes. For 
example, our studies reveal that the energy of the Kohn mode, 
namely sloshing mode, decreases on steering simply connected or pancake
shaped condensate to multiply connected or toroidal condensate. This is 
attributed to the fact that in the latter, the predominant excitations are 
along the axis of the toroid. Hence, the effective wavenumber of the 
Kohn mode is lower than in pancake geometry. Furthermore at $T\neq0$ when the 
condensate is multiply connected, the maxima of the condensate and the
thermal densities tend to coincide. This, however, is not the case in 
simply connected condensates. Experimentally, it is possible to transform 
harmonic to toroidal traps~\cite{ryu_07}, by combining a harmonic potential
with a Gaussian potential. This is the scheme we have adopted in the present
work. Toroidal confining potential, however, can also be realized with
Laguerre-Gaussian (${\rm LG}_{p}^{l}$) laser beams, and were first 
examined in theoretical works \cite{wright_2000,courtial_97}. Here,
$p\geqslant0$ and $l$, are radial and azimuthal orders of the laser beam,  
respectively. Following which, toroidal condensates of atomic $^{23}$Na 
\cite{ramanathan_11}, and $^{87}$Rb~\cite{moulder_12} have been achieved 
using ${\rm LG}_0^l$ beams~\cite{curtis_03}. Furthermore, toroidal trapping 
potentials for $^{87}$Rb condensates have been realized by combining an 
RF-dressed magnetic trap with an optical potential \cite{heathcote_08} or by 
the intersection of three light beams as elucidated in Ref. \cite{marti_15}.
Our present investigation has profound experimental
implications since, the more commonly used approach to produce toroidal 
traps using ${\rm LG}_{p}^{l}$ laser beams do not have limiting case 
equivalent to a harmonic oscillator potential. For this reason, 
${\rm LG}_{p}^{l}$ laser beams are not suitable to examine the 
variation in fluctuations as a pancake shaped condensate is transformed to 
a toroidal one.

As mentioned earlier, 
the role of fluctuations in the generation of topological defects in 
toroidal BECs demands a good understanding of the quasiparticle modes.
To this end, we employ the gapless Hartree-Fock-Bogoliubov theory with 
the Popov approximation (HFB-Popov) to compute the energy eigenspectra of the 
quasiparticle excitations of these condensates and demonstrate its distinct
features. 

   The remaining of the paper consists of Sec.~\ref{theory}, which 
provides a brief explanation of the HFB-Popov formalism for interacting 
quasi-2D BEC to compute the quantum and thermal fluctuations. In the same 
section, we also provide a description of the Stochastic Gross-Pitaveskii 
(SGP) approach to incorporate thermal fluctuations in BECs. The results and 
discussions are given in Sec.~\ref{results}. In this section the evolution of 
the Bogoliubov quasiparticle energies and amplitudes based on present studies 
are presented.  We also highlight the change in the structure of quantum and 
thermal fluctuations due to the change in the geometry of the external 
confining potential. Finally, we end with conclusions highlighting the key 
findings of the present work in Sec.~\ref{concl}. 

\section{Theoretical methods}
\label{theory}
In a quasi-2D trapped BEC, the trapping frequencies of the harmonic oscillator
potential $V(x,y,z)=(1/2)m\omega_x^2(x^2 + \alpha^2y^2 + \lambda^2z^2)$ 
satisfy the condition $\omega_x,\omega_y \ll \omega_z $, and
$\hbar\omega_z\gg\mu$. Here,  $\alpha = \omega_y/\omega_x$ and $\lambda =
\omega_z/\omega_x$ are the anisotropy parameters along the transverse and
axial directions, respectively. For $\lambda \gg 1$ and $\alpha=1$ or quasi-2D
BEC, the axial degrees of freedom can be integrated out and only the 
transverse excitations contribute to the dynamics. Under the mean field 
approximation, in Cartesian coordinate system, the second quantized form of 
the grand-canonical Hamiltonian of the system is 
\begin{eqnarray}
\fl
  \hat{H} = \int\!\!\!\!\int dx dy\,\hat{\Psi}^\dagger(x,y,t)
        \bigg[-\frac{\hbar^{2}}{2m}\nabla^2_\perp
        + V(x,y)
        -\mu + \frac{U}{2}\hat{\Psi}^\dagger(x,y,t)\hat{\Psi}
        (x,y,t)\bigg]\hat{\Psi}(x,y,t).
\label{hamiltonian} 
\end{eqnarray}
Here,  $\hat{\Psi}$ and $\mu$ are the Bose field operator with BEC, and the 
chemical potential, respectively with 
$\nabla^2_\perp = \partial^2/\partial x^2 + \partial ^2/\partial y^2$. 
The inter-atomic interaction $U = 2 a\sqrt{2\pi\lambda}$ , where $a>0$ is 
the $s$-wave scattering length, 
and $m$ is the atomic mass. In the present work, to examine toroidal 
condensates vis-a-vis pancake shaped condensates, we superimpose a 
2D-Gaussian potential to the harmonic oscillator potential. The confining 
potential is thus
$V_{\rm net}(x,y) = V(x,y) + U_0e^{{-(x^2 + \alpha^2 y^2)}/2\sigma^2}$
with $U_0$ and $\sigma$ as the strength and radial width of the Gaussian 
potential, respectively. With  $\alpha=1$, $V_{\rm net}$ is 
rotationally symmetric, the symmetry is, however, broken 
when $\alpha<1$. The equipotential curves are then ellipses with semi-major 
axis along $y$-axis. Furthermore, when $U_0\gg0$, $V_{\rm net}$ is modified, 
and at higher values of $U_0$ the potential assumes the form of a doughnut 
or a toroid. 
Thus tuning $U_0$, we can examine how fluctuations and modes
evolve as the geometry of the condensate is transformed from pancake to
toroid. In further calculations, and numerical evaluations the spatial and 
temporal variables are scaled as $x/a_{\rm osc}$, $y/a_{\rm osc}$ and 
$\omega_x t$ respectively, where $a_{\rm osc} = \sqrt{\hbar/m\omega_x}$.


\subsection{Hartree-Fock-Bogoliubov-Popov formalism}
 We employ HFB-Popov approximation
\cite{griffin_96,gies_04,gies-04,gies-04a,gies_05,roy_14,roy_14a,roy_15,roy_15a}
to examine the properties of the collective excitations 
at $T=0$, and impact of $U_0$ on the variation in the density distribution 
at $T\neq0$. In this formalism, the Bose field operator $\hat\Psi$ is 
decomposed into
two parts; the $c$-field or the condensate part represented by $\phi(x, y, t)$
and the non-condensate or the fluctuation part denoted by $\tilde\psi(x,y,t)$. 
With $\hat{\Psi} = \phi + \tilde\psi$, in the mean-field domain
the generalized GP equation with the time-independent HFB-Popov
approximation is 
\begin{equation}
  \hat{h}\phi + U\left[n_{c}+2\tilde{n}\right]\phi = 0,
  \label{gpe1s}
\end{equation}
where, $\hat{h}= (-\hbar^{2}/2m)\left(\partial ^2/\partial x^2
                 +\partial ^2/\partial y^2\right) + V(x,y)-\mu$ represents 
the single-particle part of the Hamiltonian;
$n_{c}(x,y)\equiv|\phi(x,y)|^2$,
$\tilde{n}(x,y)\equiv\langle\tilde{\psi}^{\dagger}(x,y,t)
\tilde{\psi}(x,y,t)\rangle$, and $n(x,y) = n_{c}(x,y)+ \tilde{n}(x,y)$
are the local condensate, non-condensate, and total density,
respectively. Applying Bogoliubov transformation, $\tilde\psi(x,y,t)$ 
in terms of quasiparticle modes is
\begin{eqnarray}
   \tilde{\psi} &=&\sum_{j}\left[u_{j}(x,y)
    \hat{\alpha}_j(x,y) e^{-iE_{j}t/\hbar}
   - v_{j}^{*}(x,y)\hat{\alpha}_j^\dagger(x,y) e^{iE_{j}t/\hbar}
    \right]. \;\;\;\;\;
\label{ansatz}
\end{eqnarray}
Here, $\hat{\alpha}_j$ ($\hat{\alpha}_j^\dagger$) are the quasiparticle
annihilation (creation) operators and satisfy the usual Bose commutation
relations, and the subscript $j$ represents the energy eigenvalue index.
The functions $u_j$ and $v_j$ are the Bogoliubov quasiparticle amplitudes 
corresponding to the $j$th energy eigenstate. Using the above ansatz and 
definitions we arrive at the following Bogoliubov-de Gennes (BdG) equations
\begin{eqnarray}
\label{bdg1}
(\hat{h}+2Un)u_{j}-U\phi^{2}v_{j}&=&E_{j}u_{j},\nonumber\\
-(\hat{h}+2Un)v_{j}+U\phi^{*2}u_{j}&=&E_{j}v_{j}.
\end{eqnarray}
The thermal or non-condensate density $\tilde{n}$ at temperature $T$ is 
\begin{equation}
 \tilde{n}=\sum_{j}\{[|u_{j}|^2+|v_{j}|^2]N_{0}(E_j)+|v_{j}|^2\},
  \label{n_tilde}
\end{equation}
where, $\langle\hat{\alpha}_{j}^\dagger\hat{\alpha}_{j}\rangle = (e^{\beta
E_{j}}-1)^{-1}\equiv N_{0}(E_j)$ with $\beta=1/k_{\rm B} T$, is the Bose
factor of the $j$th quasiparticle state with energy $E_j$ at temperature $T$.
However, it should be emphasized that, when $T\rightarrow0$, $ N_0(E_j)$'s in 
Equation (\ref{n_tilde}) vanish. The non-condensate density is then reduced to
$ \tilde{n} = \sum\nolimits_j|v_{j}|^2$ which represent quantum fluctuations.

It must be mentioned here that the HFB-Popov approximation is gapless, and 
satisfies Hugenholtz-Pines theorem~\cite{hugenholtz_59}.
Using the theory it is possible to calculate the quasiparticle excitations,
and obtain the equilibrium density profiles of the condensate and the thermal 
densities. The essence of the HFB-Popov theory is to obtain stationary 
states as self-consistent solutions of Equations (\ref{gpe1s}) and 
(\ref{bdg1}). However, the omission of the anomalous average, which
accounts for the transfer of particles between condensate and non-condensate
clouds, in this theory leads to non-conservation of particle number. This is
in contrast to number-conserving theories where the increase in the number 
of condensate particles should imply a decrease in the number of 
non-condensate particles. In the present work, through normalization,
we maintain constant number of particles, and this translates to a change in
the chemical potential.
\begin{figure}[h]
\centering
 \includegraphics[width=6.0cm]{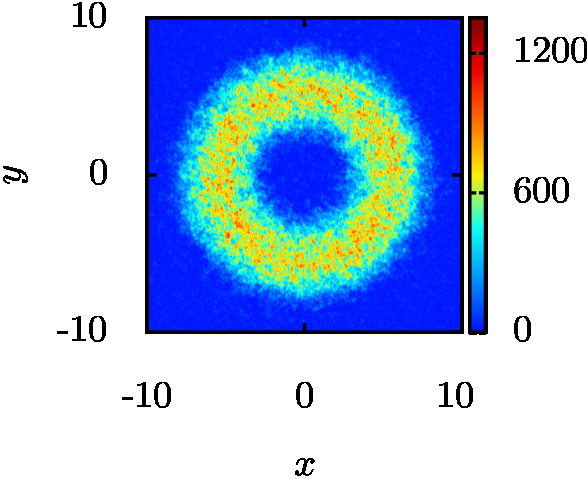}
 \caption  {(Color online)Equilibrium density profile of the $^{23}$Na
             condensate at
             $T=20$ nK with $\sim 10^5$ atoms in a toroidal trap having
             $U_0 = 45\hbar\omega_\perp$, where $\omega_\perp= 2\pi \times
             20$ Hz and $\lambda = 39.5$. This has been computed using the
             SGP method. Density is measured in units of $a_{\rm osc}^{-2}.$
             }
\label{sgp_den}
\end{figure}


\subsection{Stochastic Gross-Pitaevskii equation (SGPE)}

  The time-dependent HFB-Popov theory is well suited to study the quasiparticle
modes, and the quantum and thermal fluctuations at equilibrium. It is, 
however, not suitable to the study the dynamics of condensates~\cite{wild_11}.
The SGPE, on the other hand, provides a good description of the dynamical 
evolution of the condensate, and non-condensate clouds. In fact, the 
experimental realizations have close correspondence to the SGPE results,
where the individual results obtained with independent noise realizations are 
representatives of independent experimental observations. Hence, as one may
expect, the experimental results for a system at equilibrium are equivalent 
to the ensemble average of results from
SGPE~\cite{cockburn_09,gautam_14,cockburn_12}. For the 
present work we employ SGPE to examine the realistic density distribution of
the condensate and non-condensate clouds. This provides an analysis of the 
thermal fluctuations sans the knowledge of quasi particle spectrum, and a 
comparison with the HFB-Popov results has the potential to understand the 
effects of fluctuations. In the present work we use a scheme based on solving 
the stochastic Langevin equation for BEC~\cite{ford_88} in toroidal traps, 
and the SGPE is~\cite{cockburn_12}
\begin{equation}
i\hbar\frac{\partial \Psi(\mathbf x,t)}{\partial t}
= (1-i\gamma)\left[-\frac{\hbar^2\nabla^2}{2m}
+ V_{\rm net} +U|\Psi(\mathbf x,t)|^2-\mu\right]\Psi(\mathbf x,t)
+ \eta(\mathbf x,t).
\label{sgpe}
\end{equation}
The constant $\gamma$ denotes the dissipation, and $\mu$ is the chemical 
potential. Following the fluctuation-dissipation theorem required for 
equilibration, the random fluctuation $\eta(\mathbf x,t)$ satisfies the 
following noise-noise correlation
\begin{equation}
\langle \eta(\mathbf x,t) \eta^*(\mathbf x',t')\rangle = 2\gamma k_B T\hbar
                             \delta(\mathbf{ x-x'})\delta({t-t'}),
\end{equation}
where $k_B$ and $T$ are the Boltzmann constant and temperature respectively,
and $\langle \ldots \rangle$ denotes the ensemble average, where each member
represents realizations with difference arising from fluctuations. The 
essence of SGPE is to divide the system into two parts. In which a few 
highly occupied low-energy quasiparticles are subsumed in the Langevin field 
$\Psi({\mathbf x}, t)$, and the other part represents a heat bath 
denoted by the noise term $\eta$. The latter also takes incorporate the 
effects of the higher energy quasiparticles. Another description of the 
Langevin field is, it is  the expectation of the field operator 
$\hat{\Psi}$ at zero temperature, and at finite temperatures has 
contributions from the thermal fluctuations. 

In the SGP formalism, the physical properties are computed as an ensemble
average over several independent noise realizations, and the total number 
of atoms at equilibrium is fixed by the chemical potential $\mu$ 
irrespective of the change in temperature ~\cite{shukla_13}. As an example
the density profile or $|\Psi({\mathbf x}, t)|^2$ of one realization is 
shown in Fig. \ref{sgp_den}. From the density, at equilibrium, the 
quasicondensate density $n_{\rm qc}(x)$ is extracted through the second 
order correlation \cite{prokofev_01,cockburn-11,proukakis_06} as
\begin{equation}
 n_{\rm qc}(x) = n(x)\sqrt{2 - g^{(2)}(x)}, 
\end{equation}
where $n(x) = |\Psi(x)|^2$, and
$g^{(2)}(x) = \langle|\Psi(x)|^4\rangle/\langle|\Psi(x)|^2\rangle^{2}$ is
the second order correlation function. Thus the quantity $n(x) -
n_{\rm qc}(x)$ gives us the non-condensate or the thermal density.
Pertinent to the present work and as mentioned earlier,
thermal fluctuations in toroidal potentials have a direct bearing on the
KZ mechanism as discussed in Refs.~\cite{das_12,weiler_08}. These
previous theoretical investigations, however, regard the toroidal systems 
as quasi-1D. In the present work we examine the validity of this
assumption in the context of finite temperatures effects, and 
the observations based on the results from HFB-Popov and SGP theories
are given in the next section.


\section{Results and Discussions}
\label{results}
\begin{figure}
\centering
 \includegraphics[width=8.0cm]{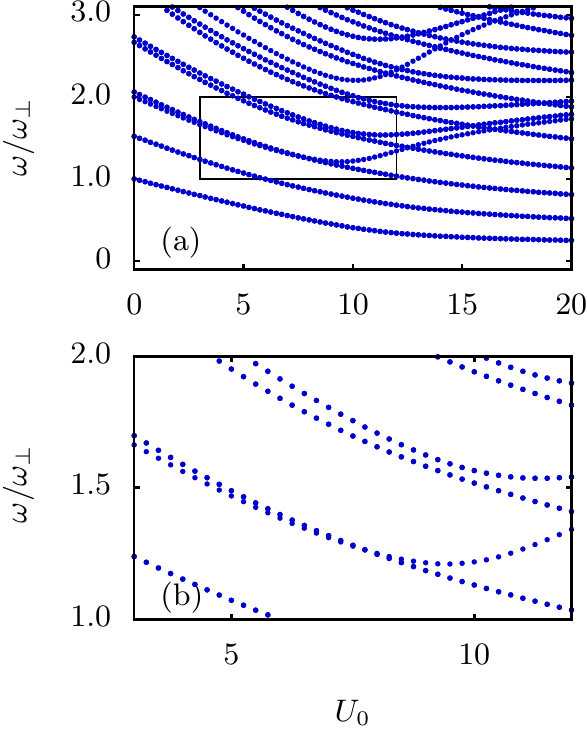}
 \caption {(Color online) The evolution of the mode energies as a function 
           of $U_0$ for
           $\alpha=1$ at $T=0$. (a) Shows the evolution of the low-lying mode
           energies in the domain $0\leqslant U_0 \leqslant 20\hbar
           \omega_\perp$
           for $N_{\rm Na} = 2\times 10^3$. (b) The enlarged view of the region
           enclosed within the black rectangular box in (a) to resolve the
           actual level-crossing and quasidegeneracy of modes.
          }
\label{mode_evol_sol}
\end{figure}

To examine the quasiparticle spectrum with our theoretical scheme, we 
take BEC of atomic $^{23}$Na, with scattering length $a = 53.3a_0$ 
\cite{davis_95,stamper_kurn_98}, as a test case. Here $a_0$
is the Bohr radius. The evolution of the
quasiparticle modes with the variation of $U_0$ is computed for 
$N_{\rm Na} = 2\times 10^3$ with
$\lambda = 39.5$, and $\omega_x=\omega_y=\omega_{\perp} = 2\pi \times 20.0$ Hz.
The evolution of the low-lying mode energies are shown in 
Figure \ref{mode_evol_sol} as a function of $U_0$. It is evident, from the 
trends in the plot, there are several transformations or changes in the
nature of the modes as the condensate is modified from pancake to toroidal
geometry. 

\subsection{Decrease in Kohn mode energy}
\label{kohn}
\begin{figure}[t]
\centering
 \includegraphics[width=9.0cm]{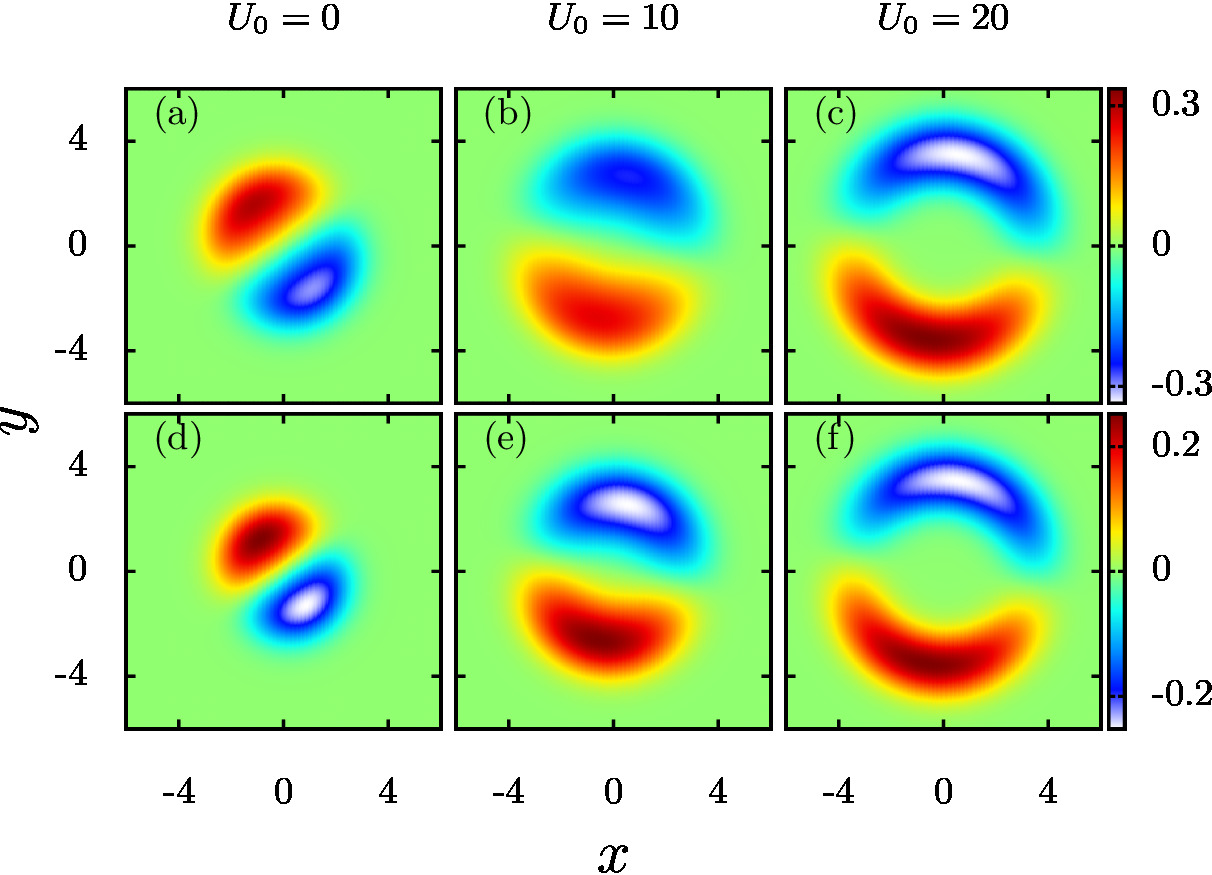}
 \caption {(Color online) Evolution of quasiparticle amplitudes corresponding 
           to the Kohn mode as $U_0$ is increased from $0$ to
           $20\hbar\omega_\perp$ for $\alpha=1$. (a) - (c) Show the $u_{\rm Na}$
           corresponding to $U_0 = 0, 10, 20 \hbar \omega_\perp$, respectively.
           Harmonic trapping potential is applicable when $U_0 = 0$, otherwise,
           it is a Mexican hat potential. At $U_0 = 20 \hbar \omega_\perp$, 
           a toroidal shaped BEC
           is formed and the Kohn modes get deformed. (d) - (f) Show the
           $v_{\rm Na}$ corresponding to $U_0 = 0, 10, 20 \hbar
           \omega_\perp$,
           respectively. In the plots $u$ and $v$ are in units of
           $a_{\rm osc}^{-1}$. Here $x$ and $y$ are measured in units
           of $a_{\rm osc}$.
          }
\label{kmode}
\end{figure}

In the absence of the Gaussian potential or when $U_0 = 0$,  the profile of
$n_c$ has rotational symmetry with the maximum located at the
origin, and decreases to zero with $r>0$. The corresponding quasiparticle
spectrum has a Goldstone mode, and doubly degenerate Kohn modes with
$\omega/\omega_\perp=1$. The quasiparticle amplitudes corresponding to one
of the degenerate Kohn modes is shown in Figure~\ref{kmode}(a).
For $U_0 \neq 0$, $n_c$ shows a dip in the central region, and 
an overall increase in the radial extent due to the repulsive Gaussian 
potential. As $U_0$ is increased when $U_0\approx \mu$, which occurs when 
$U_0\approx15\hbar\omega_{\perp}$, the condensate cloud is toroidal in shape
and hence $n_c(0,0)\rightarrow 0$. 
The change in the geometry, pancake to toroidal, of the $n_c$ modifies the 
excitation spectrum and the structure of the Bogoliubov quasiparticle 
amplitudes. An important observation associated with the change is, the 
wavelength of excitations becomes longer as they now lie along the 
circumference of the toroid. This decreases the quasiparticle energies,
and the evolution of mode energies as a function 
of $U_0$ is shown in Figure~\ref{mode_evol_sol}. From the figure, the Kohn 
mode getting soft with increase in $U_0$ is discernible. The metamorphosis 
of the Kohn mode amplitude from pancake to toroidal geometry of the BEC 
is shown in 
Figure~\ref{kmode}. From the mode functions the ratio of the Kohn 
mode wavelength at $U_0=20\hbar\omega_\perp$ and 
$U_0=0$ is $\approx 4.0$, which is close to $\pi$, the ratio between the 
diameter and circumference of a circle. It deserves to be
mentioned here that for $15\hbar\omega_\perp < U_0 < 20\hbar\omega_\perp$, 
the energy of the
Kohn mode gets saturated. This is because the variation in the 
circumference of the condensate is small, and the change in the wavelength
of excitations is negligible. 

 One distinct trend in the evolution of selected neighbouring mode 
energy pairs is discernible from the figure. As an example, consider the 
evolution of breathing and hexapole modes with $\omega/\omega_\perp = 2.00$, 
and $2.06$, respectively, at $U_0=0$. As $U_0$ is increased, in the domain
$0< U_0 <7.5\hbar\omega_\perp$ energies of these two modes are lowered
with decreasing separation. At $ U_0\approx <7.5\hbar\omega_\perp$ 
the two modes cross over, and when $U_0>7.5\hbar\omega_\perp$ the breathing 
mode energy increases, however, the hexapole mode energy continues to decrease.
The reason behind the increase in breathing mode energy is because it is a 
$l=0$ mode it remains a radial excitation. So, 
when $U_0>7.5\hbar\omega_\perp$ as the trapping potential assumes toroidal 
geometry, the effective radial frequency is $\approx1.63\omega_\perp$.
The hexapole mode, which has $l=3$ is transformed to 
one along axis of the toroid, and hence, decreases in energy. Similar
trend occurs in other neighbouring pairs of states with $l=0$, and the other
with $l>0$. It is to be mentioned that, the modes cross over. These crossing 
are not avoided because $l$ is a good quantum number, and quasiparticles 
with different $l$ do not mix.


\subsection{Quantum and thermal fluctuations}
\label{qft}
   At zero temperature the quantum fluctuations, which depend on the $|v|^2$ 
of the quasiparticle amplitudes, produce finite $\tilde{n}$. 
Upon closer inspection, we find that the 
dominant contribution to the number of non-condensate atoms 
$\tilde{N} = \int \tilde{n}(x,y) \,dxdy$ arises from the doubly degenerate 
Kohn mode. With the variation of $U_0$, there is an enhancement of $\tilde{n}$
as $U_0$ is increased. There are two reasons behind this, and both are 
associated with the transition of $n_c$ from pancake to toroidal
structure. First, the Kohn mode energy is lowered, and second, the extrema of
the mode functions coincide with the density maxima of $n_c$.

\begin{figure}[h]
\centering
 \includegraphics[width=11.0cm]{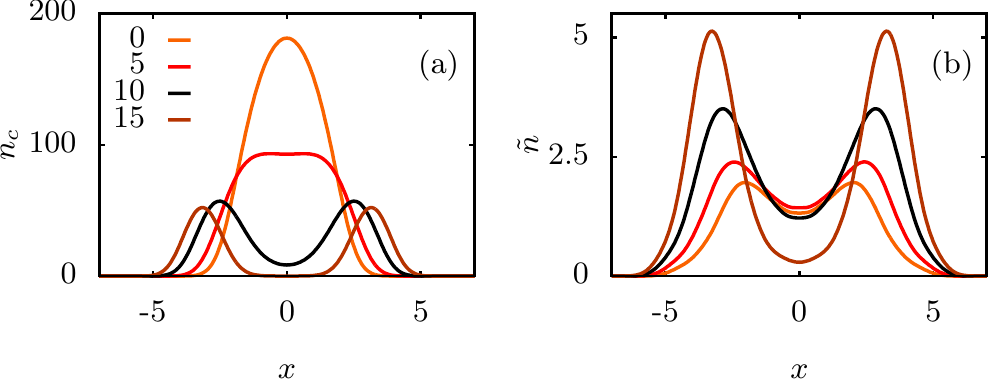}
 \caption  {(Color online) The plots along $x$-axis at $y=0$ showing the 
            variation in 
           (a) condensate and (b) non-condensate density profiles at $T=10$ nK 
           for different values of $U_0$. In the plots $n_c$ and
            $\tilde{n}$ 
           are in units of $a_{\rm osc}^{-2}$. 
           Here $x$ and $U_0$ are 
           measured in units of $a_{\rm osc}$ and $\hbar\omega_{\perp}$,
           respectively.}
\label{den}
\end{figure}

\begin{figure}[t]
\centering
 \includegraphics[width=10.5cm]{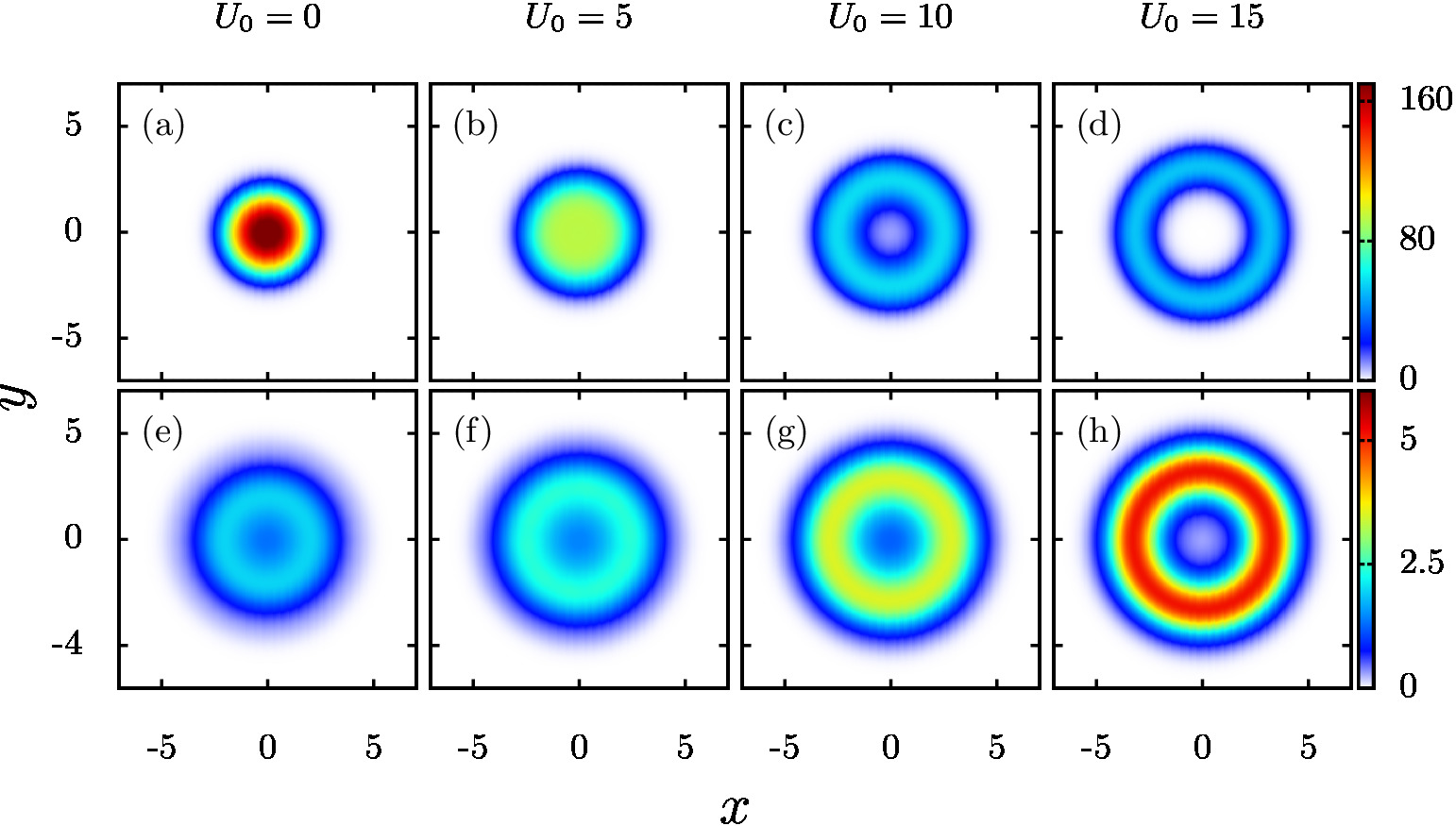}
 \caption  {(Color online) Condensate and non-condensate density plots at 
            $T=10$ nK for
           different values of $U_0$. (a) - (d)  plots of condensate, and 
           (e) - (f) non-condensate density distribution for 
           $U_0 = 0, 5, 10, 15\hbar\omega_\perp$, respectively. 
           Here $x$ and $y$ are measured 
           in units of $a_{\rm osc}$.}
\label{dentemp}
\end{figure}

 For $T\neq0$,  the Bose factor $N_0\neq0$, so in addition to the quantum 
fluctuations $\tilde{n}$ have contributions from the thermal fluctuations 
as well. The latter, however, are orders of magnitude larger in 
the temperature domain of experimental interest $\approx 10$ nK. Albeit
we have done computations for different temperatures, for the aforementioned
reason, we examine the HFB-Popov results for $T=10$ nK. The higher
$\tilde{n}$ at finite temperature enhances the interaction between the 
condensate and non-condensate atoms, represented by the $\tilde{n}\phi$ term 
in Equation~(\ref{gpe1s}), and modifies $n_{c}$. Due to the repulsive 
inter-atomic interactions the profile of $\tilde{n}$ develops a dip in the 
central region when $U_0=0$, where the maximum of $n_c$ is located. In 
general, as evident from the density plots in Figure~\ref{den}, due to higher 
energy the spatial extent of the thermal cloud is larger than the condensate 
cloud.
With higher $U_0$, $n_{c}$ develops a dip at the center, and 
as mentioned in the previous section, is transformed to a toroidal condensate 
when $U_0\approx15\hbar\omega_\perp$. The transformation is
evident from the density plots of $n_c$ as shown in Figure~\ref{dentemp}. As
$n_c$ and $\tilde{n}$ are coupled, the density structure of $\tilde{n}$
is also altered. At $U_0=15\hbar\omega_\perp$, $n_c(0,0)\approx 0.0$ where as 
$\tilde{n}(0,0)$ is still $\approx 6\%$ of the maximum value. 
This difference is due to lower $n_c$ in the
central region, and higher energy of the non-condensate atoms. The progressive
depletion of $n_c$ from the central region with higher $U_0$, and the
corresponding changes in $\tilde{n}$ are prominent in 
Figure~\ref{dentemp} (d), (h).  At lower amplitude of the Gaussian 
potential, $U_0<15\hbar\omega_\perp$, the maxima of $n_c$ 
and $\tilde{n}$ do not coincide 
and the separation is $\approx 2.0a_{\rm osc}$. However, when
$U_0=15\hbar\omega_\perp$ the
separation between maxima is reduced to $\approx 0.1a_{\rm osc}$,
the trend in the shifting of the maxima is shown in Figure~\ref{den}
through the plot of densities along the $x$-axis. The number of thermal 
atoms increases $\approx 2.5$ fold with increasing the strength 
of $U_0$ from $0$ to $15\hbar\omega_\perp$.


\subsection{KZ scaling and fluctuations}

  The KZ mechanism has been the subject of intense investigations
in toroidal or multiply connected condensates~\cite{weiler_08,das_12,su_13}. 
One key simplifying assumption in all the works is the consideration of a 
toroidal condensate as a 1D system with periodic boundary condition. A natural 
outcome of this assumption is, at finite temperatures the maxima of the 
condensate and non-condensate densities must lie along the axis of the toroid. 
Hence, the KZ scaling laws obtained in the previous 
works ~\cite{weiler_08,das_12,su_13} are applicable in toroidal condensates
only when the parameters of the toroid satisfies the conditions 
$\mu\ll U_0$, and this is an additional criterion to ensure
the alignment of the thermal fluctuations. Otherwise, as observed in the 
present results, the non-condensate or fluctuations distribution have a 
different geometry, and the system deviates from the 1D approximation. This 
is an important finding which can only be obtained from finite temperature 
theories.

\section{Conclusions}
\label{concl}
 There is a decrease in the Kohn and other mode energies at $T=0$ as the
external trapping potential is changed from harmonic to a toroidal trapping
potential. However, the trend is not general; close to the pancake to toroidal
condensate transition, energies of all the modes with $l=0$ increase.
Our results provide an understanding on why the nature of fluctuations in
toroidal condensates are different from the usual harmonic potential trap.
Apart from this, in toroidal traps,
the breathing and the hexapole mode contribute to a distinct deviation
in the nature of fluctuations from the harmonic trapping potential.
This change is evident for $T=10$nK, when the condensate and thermal 
densities tend to coincide in toroidal traps. Most important, for the 
form of the potential considered, combination of a Gaussian and harmonic
trapping potentials, the criterion $\mu\ll U_0$ is a necessary condition to
treat the toroidal potential as quasi-1D. This condition highlights the 
different domains in the geometries of thermal fluctuations in toroidal 
condensates, and hence, the KZ scaling is applicable only when this criterion
is satisfied. Otherwise, the simplifying assumption of quasi-1D geometry
for toroidal condensates is not applicable.


\ack
We thank K. Suthar, S. Bandyopadhyay and R. Bai for useful
discussions. We acknowledge valuable discussions with
S. Gautam on SGP method. We also thank the anonymous referees for their 
thorough review and valuable comments, which contributed to improving the 
quality of the manuscript. The results presented in the paper are based on the
computations using Vikram-100, the 100TFLOP HPC Cluster at Physical Research 
Laboratory, Ahmedabad, India.

\bibliography{/home/arko/Documents/tex/main_bib_entry/bec}{}
\end{document}